\documentclass[a4paper,twocolumn,prl,showpacs]{revtex4}
\usepackage{graphicx,amsmath,amssymb}
\renewcommand{\[}{\begin{equation}}
\renewcommand{\]}{\end{equation}}
\def\bea{\begin{eqnarray}}
\def\eea{\end{eqnarray}}
\def\nn{\nonumber\\}

\newcommand{\intr}{\int d{\bf r} \;}
\newcommand{\emi}[1]{{\rm e}^{-i #1}}
\newcommand{\ei}[1]{{\rm e}^{i #1}}

\newcommand{\CP}{{\cal P}}
\newcommand{\CH}{{\cal H}}
\newcommand{\OO}{{\cal O}}
\newcommand{\FF}{{\cal F}}
\newcommand{\GG}{{\cal G}}
\newcommand{\vc}{V_{\rm cell}}

\newcommand{\p}{{\bf p}}
\newcommand{\q}{{\bf k}}
\renewcommand{\k}{{\bf k}}
\newcommand{\kk}{\mbox{\boldmath$\kappa$}}

\renewcommand{\v}{{\bf v}}
\renewcommand{\r}{{\bf r}}

\newcommand{\da}{\partial_{k_\alpha}}
\newcommand{\db}{\partial_{k_\beta}}
\newcommand{\dka}{\partial_{\kappa_\alpha}}
\newcommand{\dkb}{\partial_{\kappa_\beta}}

\newcommand{\dk}{[d \, {\rm k}]}
\newcommand{\intk}{\int_{\rm BZ} \!\!\!\! [d \, {\rm k}] \;}

\newcommand{\equ}[1]{Eq.~(\ref{#1})}

\def\bra#1{\langle#1\vert}
\def\ket#1{\vert#1\rangle}
\def\ev#1{\langle#1\rangle}
\def\me#1#2#3{\langle#1| \, #2 \, |#3\rangle}
\def\runtime{(\the\time)\qquad\the\month/\the\day/\the\year}% get current time
\def\today
 {\count10=\year\advance\count10 by -2000 \number\day--\ifcase
  \month \or Jan\or Feb\or Mar\or Apr\or May\or Jun\or
             Jul\or Aug\or Sep\or Oct\or Nov\or Dec\fi--\number\count10}

\def\hour{\count10=\time\count11=\count10
\divide\count10 by 60 \count12=\count10
\multiply\count12 by 60 \advance\count11 by -\count12\count12=0
\number\count10 :\ifnum\count11 < 10 \number\count12\fi\number\count11}

\begin{document}

%%%%%%%%%%%%%%%%%%%%%%%%%%%%%%%%%%%%%%%%%
\title{Geometrical meaning of the Drude weight \\ and its relationship to orbital magnetization}

\author{Raffaele Resta}

\affiliation{Dipartimento di Fisica, Universit\`a di Trieste, Strada Costiera 11, 34151 Trieste, Italy,}

\affiliation{Consiglio Nazionale delle Ricerche (CNR)ÐIstituto Officina dei Materiali (IOM) DEMOCRITOS, 34136 Trieste, Italy}

\affiliation{Donostia International Physics Center, 20018 San Sebasti\'an, Spain}

\date{\today}

\begin{abstract}
At the mean-field level the Drude weight is the Fermi-volume integral of the effective inverse mass tensor. I show here that the deviation of the inverse mass from its free-electron value is the real symmetric part of a geometrical tensor, which is naturally endowed with an imaginary antisymmetric part. The Fermi-volume integral of the latter yields the orbital magnetization. The novel geometrical tensor has a very compact form, and looks like a close relative of the familiar metric-curvature tensor. The Fermi-volume integral of each of the two tensors provides (via real and imaginary parts) a couple of macroscopic observables of the electronic ground-state. I discuss the whole quartet, for both insulating and metallic crystals.
\end{abstract}

\date{run through \LaTeX\ on \today\ at \hour}


\maketitle

The geometry of the occupied Bloch manifold in crystalline systems is at the root of several intensive observables of the electronic ground state. These observables obtain---in both the metallic and insulating case---as Fermi-volume integrals of the appropriate geometrical quantity. Foremost among them, and by now rather popular, is the metric-curvature tensor \cite{Provost80,Marzari97,Souza00,rap_a31,Marzari12} of the occupied Bloch manifold $\FF_{\alpha\beta}(\k)$, for which I adopt the compact expression of \equ{FF} below. Its antisymmetric imaginary part (times $-2$) is the Berry curvature, and is known to yield the anomalous Hall conductivity (AHC) \cite{Nagaosa10} in both insulators and metals; in the latter case extrinsic contributions are also necessarily present besides the intrinsic geometrical one. The real symmetric part of $\FF_{\alpha\beta}(\k)$ provides an observable as well (discussed below), although in the insulating case only.

The main formula of the modern theory of orbital magnetization \cite{Xiao05,rap128,rap130,Souza08} is the Fermi-volume integral of the antisymmetric imaginary part of a Bloch tensor, looking like a kind of modification of the Berry curvature. The existing proofs of the formula start from the textbook definition: orbital moment of a bounded sample, divided by the sample volume. This quantity is then transformed into something which makes sense even for an unbounded sample \cite{Souza08,rap148,rap150,rap151}, and eventually cast as a $\k$-space integral. It was not clear what the corresponding real symmetric tensor would be, for at least a couple of reasons: the proofs are ``antisymmetric'' since the beginning (they start from $\r \times \v$); furthermore real symmetric terms are discarded at some intermediate steps. Here I find the complete tensor (real and imaginary parts, no terms discarded), and I cast this novel geometrical tensor $\GG_{\alpha\beta}(\k)$ in a very compact form: \equ{trace} below. When integrated over the Fermi volume, its imaginary part yields magnetization, while its real part yields the Drude weight; more precisely, the crystalline correction to the free-particle Drude weight, whose geometrical meaning is made here perspicuous. The result is nontrivial for metals only. The two geometrical tensors $\FF_{\alpha\beta}(\k)$ and $\GG_{\alpha\beta}(\k)$ provide therefore, via their real and imaginary parts, a quartet of ground-state physical observables: for the sake of completeness I discuss here in some detail even the known observables which stem from $\FF_{\alpha\beta}(\k)$. In the final part of this Letter I abandon lattice periodicity and I address the same quartet of observables for a bounded sample, where the single-particle orbitals are square-integrable. 

In a crystalline solid the mean-field single-particle orbitals have the Bloch form $\ket{\psi_{j \k}} = \ei{\k \cdot \r} \ket{u_{j \k}}$; here they are normalized to one over the unit cell. The periodic orbitals $\ket{u_{j \k}}$ are eigenstates of $\CH_\k = \emi{\k \cdot \r} \CH \ei{\k \cdot \r}$, and we choose a gauge which makes them smooth in the whole Brillouin zone (BZ). I get rid of trivial factors of two throughout, thus addressing ``spinless electrons'' in all formulas. Equivalently, all formulas are given per spin channel.

Electronic ground-state properties obtain from the Bloch projector \[ \CP_{\k} = \sum_j \theta(\mu - \epsilon_{j\q}) \ket{u_{j\q}} \bra{u_{j\q}}, \] where $\mu$ is the Fermi level; gauge-invariant geometrical quantities obtain from $\k$-derivatives of $\CP_{\k}$: \bea &&\da \CP_{\k} = - \sum_j \delta(\mu - \epsilon_{j\k} ) \da \epsilon_{j\k} \;  \ket{u_{j\q}} \bra{u_{j\q}} \label{singular} \\ &+& \sum_j \theta(\mu - \epsilon_{j\k} ) ( \ket{u_{j\q}} \bra{\da u_{j\q}} +  \ket{\da u_{j\q}} \bra{ u_{j\q}} ) \nonumber  . \eea The  $\delta$-like singularity at the Fermi surface vanishes in insulators; as for the remaining contributions, they are smooth in insulators and piecewise continuous in metals. 

The main geometrical ingredients of the present work will be products of the form $(\da \CP_{\k})(\db \CP_{\k})$: they are smooth in insulators and singular in metals; nonetheless the $\delta$-singularity affects the symmetric part only, while the antisymmetric one is piecewise continuous. At this point it is expedient to provide a useful lemma. Let $\OO_\k$ be any operator diagonal on the $\ket{u_{j \k}}$ basis, with eigenvalues $o_{j\k}$. Neglecting for the time being the $\delta$-singular term the lemma states that 
\begin{widetext} \bea \mbox{Tr } \{ \OO_\k (  \da\!\! \CP_\k )( \db\!\! \CP_\k ) \}  &=&   \sum_{j} \theta(\mu - \epsilon_{j\k})  o_{j\k} \ev{\da u_{j\q} | \db u_{j\q}} + \sum_{j} \theta(\mu - \epsilon_{j\k}) \me{\db u_{j\q}}{\OO_\k ({\cal I} - 2 \,\CP_\k)}{\da u_{j\q}}  ; \label{lemma} \eea \end{widetext} this is proved via a straightforward although somewhat tedious calculation. 

I start with the special case where $\OO_\k = \CP_\k$ and I define \[ \FF_{\alpha\beta}(\k) = \mbox{Tr } \{ \CP_\k (  \da\!\! \CP_\k )( \db\!\! \CP_\k ) \}  . \label{FF} \] From \equ{lemma} it is easy to verify that  $\FF_{\alpha\beta}(\k)$ is the familiar metric-curvature tensor \cite{Marzari97,Souza00,rap_a31,Marzari12}, defined for the insulating case only. Indeed, one of the common forms for the Berry curvature of the occupied manifold is $i \mbox{Tr } \{ \CP_\k [ \da\!\! \CP_\k , \db\!\! \CP_\k ] \} $; this is well defined even for metals, where it is piecewise smooth and integrable. Instead the symmetric part of $\FF_{\alpha\beta}(\k)$ has a $\delta$-like singularity in metals.

The main object of the present Letter is another geometrical tensor, which also may be expressed according to \equ{lemma}. All crystalline ground-state properties are Fermi-volume integrals, where the integrand is a function of $\CH_\k - \mu$ \cite{nota}; in fact
$\CP_{\k}=\theta(\mu - \CH_\k)$. Here I choose to identify $\OO_\k$ with $|\CH_\k - \mu|$: \[ \GG_{\alpha\beta}(\k) = \mbox{Tr } \{ |\CH_\k - \mu| (  \da\!\! \CP_\k )( \db\!\! \CP_\k ) \} . \label{trace} \] By definition \[ |\CH_\k - \mu| = (\CH_\k - \mu) ({\cal I} - 2 \,\CP_\k); \] since this operator vanishes at the Fermi level, it annihilates the $\delta$-like singularity in $(\da \CP_{\k})(\db \CP_{\k})$. Therefore $\GG_{\alpha\beta}(\k)$ is integrable (both symmetric and antisymmetric parts) even in metals.
Given that $({\cal I} - 2 \,\CP_\k)^2 = {\cal I}$, the lemma of \equ{lemma} yields:
\begin{widetext} \bea  \GG_{\alpha\beta}(\k)  &=&   \sum_{j} \theta(\mu - \epsilon_{j\k})  (\mu - \epsilon_{j\k} )\ev{\da u_{j\q} | \db u_{j\q}}  + \sum_{j} \theta(\mu - \epsilon_{j\k}) \me{\db u_{j\q}}{( \CH_\k - \mu)}{\da u_{j\q}} , \label{GG} \\  \mbox{Re } \GG_{\alpha\beta}(\k) &=&  \sum_{j} \theta(\mu - \epsilon_{j\k}) \; \mbox{Re } \me{\da u_{j\q}}{( \CH_\k - \epsilon_{j\k})}{\db u_{j\q}} , \label{GG1} \\ \mbox{Im } \GG_{\alpha\beta}(\k) &=& \sum_{j} \theta(\mu - \epsilon_{j\k}) \; \mbox{Im } \me{\da u_{j\q}}{(2\mu - \CH_\k - \epsilon_{j\k})}{\db u_{j\q}} . \label{GG2}\eea \end{widetext}
I draw attention to the fact  that $\GG_{\alpha\beta}(\k)$ defined as a trace, \equ{trace}, is gauge-invariant in the generalized Marzari-Vanderbilt sense \cite{Marzari97,Marzari12}, i.e. it is invariant for unitary transformations of the occupied $\ket{u_{j\q}}$, smooth in $\k$. Instead the above expression for $\GG_{\alpha\beta}(\k)$ as a sum over states requires the ``Hamiltonian gauge'', i.e. the $\ket{u_{j\q}}$ are actual eigenstates of $\CH_\k$.

At this point it is easy to recognize in \equ{GG2} the main formula of the modern theory of orbital magnetization \cite{rap130,Souza08,rap_a30}. In fact \[ M_\gamma = - \frac{e}{2 \hbar c} \varepsilon_{\gamma\alpha\beta} \intk  \mbox{Im } \GG_{\alpha\beta}(\k) , \label{magne} \] where $\varepsilon_{\gamma\alpha\beta}$ is the antisymmetric tensor, the sum over repeated indices is implicit, and $\dk = d\k/(2\pi)^d$ ($d$ is the dimension, either 2 or 3). The BZ integration here---as well as in the following equations---is actually a Fermi-volume integration, owing to the Fermi $\theta$ function in the definition of $\FF_{\alpha\beta}(\k)$ and $\GG_{\alpha\beta}(\k)$.

The real symmetric part of $\GG_{\alpha\beta}(\k)$ provides, as anticipated, a geometrical formulation for the Drude weight. In a crystalline metal the intrinsic part of the conductivity tensor is \[ \sigma_{\alpha\beta}(\omega) = D_{\alpha\beta} \left[ \delta(\omega) + \frac{i}{\pi \omega} \right] +\sigma_{\alpha\beta}^{(\rm reg)}(\omega) . \] The regular (interband) contribution $\sigma_{\alpha\beta}^{(\rm reg)}(\omega)$ is a linear-response property, while the Drude weight $D_{\alpha\beta}$ is a ground-state property. One of its mean-field expressions is \cite{Allen06}: \[ D_{\alpha\beta} = \pi e^2 \sum_{j} \intk \theta(\mu - \epsilon_{j\k}) \, m^{-1}_{j,\alpha\beta}(\k) , \] where the effective inverse  mass tensor of band $j$ is  \[ m^{-1}_{j,\alpha\beta}(\k) = \frac{1}{\hbar^2} \frac{\partial^2 \epsilon_{j\k}}{\partial k_\alpha \partial k_\beta} . \] $D_{\alpha\beta}$ vanishes in insulators, while in metals it can be transformed into a Fermi-surface integral: it acquires then the meaning of an intraband term \cite{Allen06}.

Starting from the identity $\me{u_{j\q}}{( \CH_\k - \epsilon_{j\k})}{u_{j\q}} \equiv 0$ and taking two derivatives, one arrives at \bea  m^{-1}_{j,\alpha\beta}(\k) &=& \frac{1}{m} \delta_{\alpha\beta} - \frac{2}{\hbar^2} \mbox{Re } \me{\da u_{j\q}}{( \CH_\k - \epsilon_{j\k})}{\db u_{j\q}} , \nn D_{\alpha\beta} &=& \pi e^2  \frac{n}{m}\delta_{\alpha\beta} - \frac{2\pi e^2}{\hbar^2}  \intk \mbox{Re } \GG_{\alpha\beta}(\k)  , \eea where $n$ is the electron density. The first term on the r.h.s. is the free-electron Drude weight, while the second one is the geometrical correction due to the crystalline potential; the relationship to \equ{magne} is perspicuous.

I give now more details about the physical meaning of the observables obtained as Fermi-volume integrals of the two other geometrical quantities cited above: real and imaginary parts of the metric-curvature tensor $\FF_{\alpha\beta}(\k)$, \equ{FF}. As for $\mbox{Re } \FF_{\alpha\beta}(\k)$, it is a regular expressions only in insulators; its Cartesian trace is then related to the Marzari-Vanderbilt gauge-invariant quadratic spread \cite{Marzari97,Marzari12} indicated as $\Omega_{\rm I}$ in the literature:  \[ \Omega_{\rm I} =  \vc  \intk  \mbox{Re } \FF_{\alpha\alpha}(\k) , \] where $\vc$ is the cell volume (area for $d=2$). Although introduced as a formal quantity, $\Omega_{\rm I}/\vc$ is a genuine physical observable. In fact, it has been shown in 2000 by Souza, Wilkens, and Martin \cite{Souza00} that $\Omega_{\rm I}/\vc$ is related to longitudinal conductivity via the sum rule \[ \frac{\Omega_{\rm I}}{\vc}  = \frac{\hbar}{\pi e^2} \int_{\epsilon_{\rm g}/\hbar}^\infty \frac{d \omega}{\omega} \sum_\alpha \mbox{Re } \sigma_{\alpha\alpha} (\omega) \label{swm} , \] where $\epsilon_{\rm g}$ is the band gap of the insulating crystal.
In the metallic case one may use only the regular interband term $\sigma_{\alpha\alpha}^{(\rm reg)}(\omega)$ into \equ{swm}: the lhs member is then equal to the Fermi-volume integral of the Cartesian trace of the quantum metric, where the singular ($\delta$-like) term is omitted.

The imaginary part of $\FF_{\alpha\beta}(\k)$ (times $-2$) is the Berry curvature of the occupied manifold; as observed above it is smooth in insulators and piecewise continuous in metals. The AHC, i.e. the Hall conductivity in zero magnetic field, can be nonvanishing only if the Hamiltonian lacks time-reversal symmetry. When expressed in $e^2/h$ units (a.k.a. klitzing$^{-1}$) it is dimensionless for $d=2$, while it has the dimensions of an inverse length for $d=3$. The known expression for both metals and insulators is \[ \sigma_{\alpha\beta} = \frac{2 e^2}{\hbar} \intk  \;\mbox{Im } \FF_{\alpha\beta}(\k), \label{ss} \] and this expressions holds for both $d=2$ and $d=3$. In the insulating case the AHC is quantized, while in the metallic case \equ{ss} is only the intrinsic (or geometric) contribution to the AHC; other contributions, known as ``skew scattering'' and ``side jump'' must be added \cite{Nagaosa10}.

This concludes the discussion of the four geometrical observables of the electronic ground state, all of them obtained as Fermi-volume integrals of the real (symmetric) and imaginary (antisymmetric) parts of the tensors $\FF_{\alpha\beta}(\k)$ and $\GG_{\alpha\beta}(\k)$. 

From now on, I abandon lattice periodicity and I consider instead a bounded sample (possibly noncrystalline), where the single-particle orbitals $\ket{\varphi_j}$ are square-integrable: so-called ``open'' boundary conditions (OBCs). The mean-field Hamiltonian is written as \[ \CH_{\kk} = \frac{1}{2 m} (\p + \hbar \kk)^2 + V(\r) , \] where setting $\kk \neq 0$ amounts to a trivial gauge transformation, easily ``gauged away'' within OBCs. The $\kk$-dependent orbitals are in fact $\ket{\varphi_{j\kk}} = \emi{\kk \cdot \r}\ket{\varphi_j}$, and the ground state projector in Schr\"odinger representation is \[ \me{\r}{\CP_{\kk}}{\r'} = \ei{\kk \cdot (\r' - \r)} \me{\r}{\CP}{\r'} , \] where $\CP$ is the zero-$\kk$ ground-state projector: \[ \CP = \sum_{\epsilon_j \le \mu}\ket{\varphi_j}\bra{\varphi_j} .\]

Gauge-invariant geometric quantities within OBCs obtain from  $\kk$-derivatives of $\CP_{\kk}$ evaluated at $\kk =0$. i.e  \[ \me{\r}{\dka \CP_{\kk}}{\r'} = i (r_\alpha' - r_\alpha) \me{\r}{\CP}{\r'} , \] or in operator notation $\dka \CP_{\kk} = - i [ r_\alpha, \CP ]$. This operator encodes the linear response of the ground state to an infinitesimal gauge potential.

The generic operator product is then \[ (\dka \CP_{\kk}) ( \dkb \CP_{\kk}) = - [ r_\alpha, \CP ] \, [ r_\beta, \CP ] ,  \] and the analogue of the integrated metric-curvature tensor is  \bea \FF_{\alpha\beta} &=& - \frac{1}{V} \mbox{Tr } \{ \CP [ r_\alpha, \CP ] \, [ r_\beta, \CP ] \} \nn &=& \frac{1}{V} \mbox{Tr } \{ \CP r_\alpha r_\beta \CP  \} - \frac{1}{V}\mbox{Tr } \{ \CP \alpha \CP  r_\beta \CP  \} , \eea  i.e. the second cumulant moment of the electron distribution (or quantum fluctuation of the dipole). I have divided by the volume (area for $d=2$) in order to define an intensive quantity. This trace is obviously real symmetric even in absence of time-reversal symmetry; it provides the OBCs analogue of $\Omega_{\rm I}/\vc$, as discussed e.g. in Ref. \cite{rap118}. In the large-$V$ limit this quantity converges to a finite limit in insulators, and diverges in metals \cite{rap132}.

We may write the trace of $\FF_{\alpha\beta}$ in the Schr\"odinger representation, i.e. \[ \FF_{\alpha\beta} = - \frac{1}{V} \intr \me{\r}{\CP [ r_\alpha, \CP ] \, [ r_\beta, \CP ]}{\r}  ;  \] it has been shown in Ref. \cite{rap146} that the function $\mbox{Im } \me{\r}{\CP [ r_\alpha, \CP ] \, [ r_\beta, \CP ]}{\r}$ carries the information needed to extract the value of the AHC even from an OBCs calculation. It is enough to evaluate the trace per unit volume by integrating over an inner region of the sample (and {\it not} over the whole sample). For instance for a bounded crystallite \[ \sigma_{\alpha\beta} = - \frac{2 e^2}{ \hbar \vc} \int_{\rm cell} \!\!\!\! d \r \; \mbox{Im } \me{\r}{\CP [ r_\alpha, \CP ] \, [ r_\beta, \CP ]}{\r}, \label{ss2} \]  where the cell  is in the center of the crystallite, and the large-crystallite limit is taken. This is demonstrated in Ref. \cite{rap146} for insulators and in Ref. \cite{unpub} for metals.

Next I address the Drude-magnetization tensor $\GG_{\alpha\beta}$, which is based within OBCs on the operator \[ | \CH - \mu| \, (\dka \CP_{\kk}) ( \dkb \CP_{\kk}) =  -| \CH - \mu| \, [ r_\alpha, \CP ] \, [ r_\beta, \CP ] . \label{oper} \] Its imaginary part has been previously addressed in the literature, where it was dealt with in a similar way as for the AHC. One gets the magnetization by taking the trace per unit volume of $\mbox{Im } \me{\r}{| \CH - \mu| \; [ r_\alpha, \CP ] \, [ r_\beta, \CP ]}{\r'}$, using the bulk region of the sample only. The approach has been demonstrated in Refs. \cite{rap148,rap151} for insulators and in Ref. \cite{rap150} for metals; the operators actually used therein were somewhat different from \equ{oper}, but equivalent to it after the imaginary part is taken. The form of \equ{oper} was proposed in 2013 by Schulz-Baldes and Teufel \cite{Schulz13}. Finally, by analogy, I conjecture that the real symmetric part of the trace over the whole sample of \equ{oper} would provide a definition of the Drude weight implementable---at least in principle---over a bounded sample within OBCs.

I summarize the present findings together with some previously known ones as follows. Two Bloch tensors address the geometry of the crystalline ground state: $\FF_{\alpha\beta}(\k)$ and $\GG_{\alpha\beta}(\k)$. Their imaginary parts provide the AHC and the orbital magnetization, respectively; this applies to both metals and insulators. As for their real parts, $\mbox{Re }\FF_{\alpha\beta}(\k)$ provides a physical observable for insulators only. Instead $\mbox{Re }\GG_{\alpha\beta}(\k)$ provides a nontrivial observable for metals only: namely, the Drude weight. The whole quartet is represented in the following table:

\medskip

\begin{center}\begin{tabular}{|l|c|c|} \hline 
 &  Real & Imaginary\\ 
 &  symmetric &  antisymmetric\\
\hline  $\FF_{\alpha\beta}(\k)$  & Quadratic spread & Anomalous \\ 
   &(insulators only)  & Hall conductivity  \\ \hline
$\GG_{\alpha\beta}(\k)$  & Drude weight & Orbital \\ 
  & (metals only)  & magnetization  \\ \hline \end{tabular} \end{center}

\medskip

In conclusion I have reached a kind of unification by showing that two observables belonging to very different chapters of electronic structure theory---orbital magnetization and Drude weight---are two components of the same geometrical quantity, once the appropriate units are accounted for.

I have discussed these topics over the years with Raffaello Bianco, Antimo Marrazzo, and Ivo Souza; their invaluable contribution is gratefully acknowledged.
Work supported by the ONR Grant No. N00014-12-1-1041.


\begin{thebibliography}{10}
\bibitem{nota}
{ All ground-state properties must be invariant by translation of the energy
  zero, with the possible exception of energy itself}.
\end{thebibliography}
\end{document}